# Photoluminescent registration of fullerite $C_{60}$ derivatives during chemical interaction with $H_2$ and $N_2$ molecules.

VICTOR Zoryansky[1, a *], PETER Zinoviev[1, b], YURI Semerenko[1, c]

[1] B.Verkin Institute for Low Temperature Physics and Engineering of the National Academy of Sciences of Ukraine, 47 Nauky Ave., 61103 Kharkiv, Ukraine

[a] zoryansky@ gmail.com, [b] zinoviev@ ilt.kharkov.ua, [c] semerenko@ilt.kharkov.ua

**Abstract.** Single crystals of $C_{60}$ saturated with molecular hydrogen and nitrogen were studied using the spectral-luminescent method of registration in the quantum counting mode at a low temperature of 20 K. Previously, it was found that the temperature limit of the adsorption crossover (the transition from the diffusion mechanism of intercalation - physisorption to chemical interaction - chemisorption) for fullerite $C_{60}$ in an $H_2$ atmosphere is about 250 °C [1], and in an $N_2$ atmosphere – 420 °C [2], respectively. At saturation temperatures above those indicated, the process of chemical interaction of impurity molecules and the fullerite matrix occurs with the formation of new compounds. Therefore, saturation was carried out at temperatures of 300 °C for hydrogen and 450 °C for nitrogen under a pressure of 30 atm. For the first time, the results of registration of photoluminescent radiation at low temperatures from new substances based on fullerite $C_{60}$ are presented. A mixture of weakly saturated fulleranes $C_{60}H_X$ was obtained by saturation of hydrogen from the gas phase at a temperature close to the sorption crossover temperature, and their low-temperature photoluminescence was identified for the first time. The analysis of the "blue" shift of the beginning of the spectrum of such radiation allowed us to more accurately attribute the obtained material to the initial segment of the fullerane series. The presence of radiation of the azafullerene dimer (biazafullerene) $(C_{59}N)_2$ in the spectra of the reaction products of $C_{60}$ with $N_2$ was also detected. It was shown that for polycrystalline samples of $C_{60}$ saturated in a nitrogen atmosphere, the characteristic luminescence of biazafullerene with a maximum at 1.53 eV determines the intensity and shape of the entire short-wave part of the emission spectrum of the complex of synthesized substances.

### Introduction

Fullerite $C_{60}$ belongs to the class of molecular crystals and, like many of them, has an unusually high ability to adsorb simple gases. Due to high symmetry and a close to spherical shape, $C_{60}$ molecules rotate freely in the nodes of the crystal lattice under normal conditions. In this case, the mutual orientation of neighboring molecules at any given time determines the efficiency of charge transport and transfer of electronic excitations. For pure $C_{60}$, this is clearly manifested in the form of an increase in the integral luminescence intensity with decreasing temperature, when the frequency decreases and the nature of mutual reorientations of neighboring molecules changes. At this time, the probability of mutual orientation of neighboring molecules increases, favorable for coherent transfer of electronic excitations to the luminescence emission centers. The presence of impurity molecules in the fcc lattice of fullerite, as well as the degree of their polarity, affects the rotational dynamics of $C_{60}$ molecules, facilitating or hindering mutual rotations. This leads to a temperature shift and blurring of the boundaries of both the orientational phase transition at a temperature of $T_c = 260$ K, below which the rotation of molecules is possible only around the spatial diagonals of the cubic lattice, and the transition to orientational glass at a temperature of $T_g = 90$ K, below which reorientations of molecules are not possible at all. Sorption of impurity molecules by fullerite $C_{60}$ from the gas phase at elevated pressure and temperature can occur in two different ways - diffusion, with filling of intermolecular voids (physisorption) and chemical, in the form of a chemical reaction with the formation of new substances (chemisorption). Depending on the degree of filling of structural voids (usually octahedral with a diameter of 4.2 Å and less often tetrahedral with a diameter of 2.2 Å), a wide range of binary solutions of different concentrations is formed. Chemical sorption leads to the synthesis of new compounds based on fullerene $C_{60}$, and, like diffusion saturation, it accelerates with increasing temperature and pressure of the impurity gas atmosphere. Both processes are well recorded by photoluminescence methods, since they greatly affect the initial photophysical properties of the matrix substance. In the first case, the photoluminescence spectra of fullerite $C_{60}$ are transformed - their shape and structure change, the intensity of the emission bands is redistributed, the quantum yield changes, and sometimes a slight shift in the energy of the spectra is observed. In the second case, completely new photoluminescence spectra are recorded, often with a significant "blue" or "red" shift in the onset of luminescence and unexpected temperature effects. It is characteristic that these changes in photophysical properties do not disappear after annealing in a vacuum, unlike those for the case of diffusion sorption. The transition from one interaction mode to another usually occurs with a change in the intercalation temperature and is conventionally called sorption crossover. Previously, the existence of such a transition for hydrogen $H_2$ [1] and nitrogen $N_2$ [2] at a saturation temperature of 250 °C and 420 °C, respectively, for a pressure of 30 atm in the saturation chamber was established by X-ray structural analysis. At temperatures below these values, the filling of intermolecular cavities of fullerite with molecular hydrogen and nitrogen occurs due to the first mechanism. These processes inevitably affect the photoluminescent properties of the two-component complexes mentioned, and our results [3] showed that the mechanism of diffusion saturation for $H_2$ differs significantly from that for $N_2$ [4-6]. At temperatures above 250 °C and 420 °C, the chemical interaction of fullerene molecules with hydrogen and nitrogen, respectively, becomes decisive, and sorption crossover is observed.

This article presents the results of a study of the spectral and luminescent properties of the molecular systems $C_{60}+H_2$ and $C_{60}+N_2$ after saturation with impurities from the gas phase in the chemical sorption mode. The photoluminescence method was used to directly record the formation of new substances whose emission spectra correspond neither to pure fullerite $C_{60}$ [7] nor to its solid interstitial solutions with $H_2$ [8] and $N_2$ [9]. The features of obtaining complexes of weakly saturated fulleranes $C_{60}H_X$ by the $C_{60}$ intercalation method are highlighted, and the low-temperature photoluminescence spectra of their mixture, recorded for the first time at 20 K, are presented. An analysis of the observed "blue" shift of the hydrofullerite spectrum is described taking into account current literature data [10, 11], and the "red" shift of the presented spectra of the polycrystalline nitrogen-containing complex is interpreted for the first time during the analysis of its chemical composition [12].

**Experimental methodology.**

High sensitivity of the luminescent detection method in the quantum counting mode allows one to register the presence of any emitting substance in extremely low concentrations. Low temperature in such experiments allows one to see the dynamics of relaxation of electronic excitations with minimal influence of phonon states. Even with the appearance of effective quenching centers for excitations (as in the case of chemical interaction of $N_2$ molecules and the $C_{60}$ matrix) at low temperatures it was possible to confidently register the photoluminescence spectrum of a combination of new substances. In the case of solid samples, such a technique of optical spectroscopy allows one to detect changes in the chemical composition even only in the surface layers of crystals.

To study the photoluminescence of the $C_{60}+H_2$ and $C_{60}+N_2$ complexes, polycrystalline samples in the form of powder with a granule size of about 0.5 mm and a purity of at least 99.9% were used. Before saturation with hydrogen and nitrogen, the base spectra of the initial $C_{60}$ material were recorded, preliminarily degassed in a dynamic vacuum of $10^{-3}$ mm Hg at a temperature of 300 °C for 48 h. This is a necessary procedure for removing residual atmospheric gases that entered the fullerite crystals during previous storage. After recording the initial spectra, close to the characteristic pure, impurity-free spectra of crystalline fullerite $C_{60}$ [7], the annealing was repeated again. All manipulations with the samples for degassing and saturation in the chemical sorption mode took place in a special setup (Fig. 1), in the working chamber of which the required pressure and temperature were ensured with sufficient accuracy during the planned sample preparation time. As already mentioned [1, 2], for hydrogen and nitrogen the temperatures of transition from diffusion to chemical sorption under a pressure of 30 atm are 250 °C and 420 °C, respectively, above which X-ray structural control showed a rapid increase in the lattice parameter for both impurities. Therefore, the saturation modes were chosen corresponding only to the chemical interaction of impurity molecules with the matrix. It should be noted that in addition to the behavior of the fullerite lattice parameter upon saturation with impurities, another indicator of the presence of chemical sorption in the matrix-impurity system is the constancy of the acquired new photophysical and structural properties after annealing in a vacuum at an elevated temperature. Degassing of samples saturated in the diffusion sorption mode leads to a virtually complete return of all changed parameters to those characteristic of pure $C_{60}$.

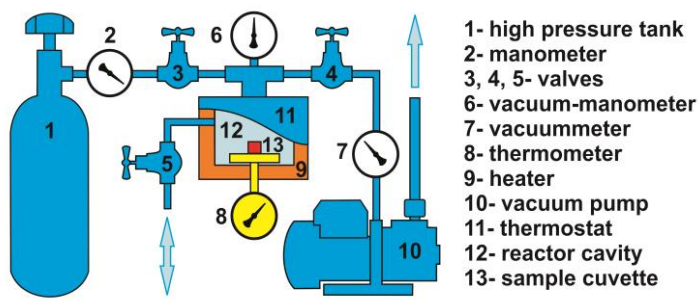

1- high pressure tank
2- manometer
3, 4, 5- valves
6- vacuum-manometer
7- vacuummeter
8- thermometer
9- heater
10- vacuum pump
11- thermostat
12- reactor cavity
13- sample cuvette

| Param. | Process | | |
|---|---|---|---|
| | Degass. | $C_{60}+H_2$ | $C_{60}+N_2$ |
| P, [atm] | $1{,}3 \cdot 10^{-6}$ | 30 | 30 |
| T, [°C] | 300-500 | 300 | 450 |
| t, [h] | 48 | 1270 | 200 |

Fig. 1. Schematic diagram of the installation for vacuum degassing of samples and their saturation in an atmosphere of the required pure gases at elevated pressure up to 70 atm. and stabilized temperature up to 600 °C.

Thus, the samples were saturated with hydrogen at 300 °C and with nitrogen at 450 °C at a gas pressure in the cell of P = 30 atm, which ensured a confident but not too intense chemical reaction. The total intercalation time of the $C_{60}$ powder for each of these gases was limited by the moment of a significant slowdown in the growth of the lattice parameter, which corresponds to impurity concentrations close to the maximum possible for these modes [1, 2].

In the case of $H_2$ saturation at a temperature 300 ºC slightly above the chemical reaction (250 ºC) threshold and a relatively low pressure, the holding time in the reactor chamber was specially chosen to be long - 1270 hours. This was done to maximize the completion of all chemical processes in the system and stabilization of the isomeric composition of the resulting composite. Under such "mild" conditions, the hydrogenation reaction of fullerene $C_{60}$ proceeds rather slowly and is generally possible mainly because the fullerene itself is its catalyst, promoting the dissociation of $H_2$ molecules. Such sample preparation conditions favor the formation of hydrofullerene or fullerane $C_{60}H_X$ molecules with a low hydrogen content, since, as a rule, the lower the reaction temperature, the lower the degree of hydrogenation. Therefore, usually, at higher temperatures, the reaction in a hydrogen atmosphere proceeds more productively, and stable forms of the fullerane series with a significant hydrogen content from $C_{60}H_{18}$ and higher are obtained relatively easily. And on the contrary, the production of weakly saturated forms, with a low hydrogenation index X from $C_{60}H_2$ to $C_{60}H_{16}$, is a separate and, as experience shows, not a simple experimental task.

Photoluminescence of $C_{60}$ crystals was recorded "in reflection" in the spectral region of 1.2–1.95 eV (1033–636 nm) with a spectral resolution of 2 nm using a high-aperture diffraction monochromator MDR-2 with an electromechanical drive and a cooled photomultiplier PEM-62 (spectral characteristic type S-1) in the photon counting mode. The cryogenic part of the experimental setup allowed changing the sample temperature in a wide temperature range of 10–230 K and stabilizing it during the experiment with an accuracy of 0.5 K. A He-Ne laser with $E_{exc}$ = 1.96 eV (632.8 nm) was used to excite photoluminescence. The excitation power density was W≤1 mW/mm$^2$. Such a limitation of the excitation power was introduced to prevent undesirable photostimulated processes in the surface layers of the studied polycrystalline samples. The luminescence measurement technique, the experimental setup, and the analytical processing of the experimental results are presented in [3].

**Results and discussion.**

It is known that the direct transition from the lowest excited electronic state (LUMO) to the highest unexcited state (HOMO) in pure $C_{60}$ fullerite is dipole-forbidden, and therefore the quantum yield of photoluminescence is weak, and its spectrum at low temperatures is a superposition of two types of radiation [7]. The first (usually more intense) is associated with the recombination of Frenkel excitons localized on defects of the crystal structure ("deep X-traps"). The second type (less intense) is due to the recombination of free singlet monomolecular Frenkel excitons with the participation of various intramolecular modes. The radiation of "deep X-traps" is luminescence that occurs as transitions between inhomogeneously broadened localized states of two neighboring molecules that are located close enough to an impurity or a strong lattice defect on which a perturbed Frenkel exciton with a decrease in symmetry is concentrated. Naturally, the intensity of $C_{60}$ luminescence critically depends on the concentration of such emission centers in crystals and the efficiency of exciton transport to these centers. Impurity molecules introduced by diffusion affect both of these circumstances, and for this reason, as was said above, the spectra of $C_{60}$-based interstitial solid solutions are most often modifications of the spectra of pure fullerite. But chemically modified $C_{60}$ fullerite, on the contrary, always has its own photoluminescence spectrum, obviously characteristic only of a newly synthesized compound or composite. As a rule, such transformations are accompanied by a strong "red" or "blue" shift of the luminescence onset, which obviously corresponds to the new energy position and structure of the HOMO and LUMO molecular orbitals of the resulting substance. It is precisely such changes that are observed in our case of chemical sorption of $H_2$ and $N_2$ from the gas atmosphere by polycrystalline $C_{60}$.

**Nitrogen chemisorption.**

Fig. 2 shows the low-temperature (20 K) photoluminescence spectra of pure polycrystalline $C_{60}$ (black solid curve in Fig. 2a) and saturated with nitrogen at 450 °C under a pressure of 30 atm (chemical sorption mode – red solid curve in Fig. 2b), corrected for the spectral sensitivity of the setup and normalized to their integral intensity. The arrow shows the "red" shift of 0.19 eV for the first recorded maximum on the short-wave side in the spectrum of modified $C_{60}$ relative to the onset of pure fullerite luminescence. This shift, taking into account some features, usually correlates with changes in the band gap in the crystals. A strong decrease in the intensity [6] of the low-temperature photoluminescence of a mixture of nitrogen-containing compounds relative to the intensity of pure $C_{60}$ was also observed in the experiment, which is not shown in Fig. 2 due to the normalization of both spectra a and b. The same result, a nearly twofold decrease in the quantum yield of fullerite chemically modified with nitrogen relative to pure $C_{60}$, is discussed in [12], devoted to the study of the luminescence of a solution of azafullerene dimers in toluene. The spectrum of biazafullerene $(C_{59}N)_2$ is shown in Fig. 2b by a dotted green line [12] (the wavelength is converted to the energy of quanta). This glow coincides well in shape and position with the first band in the short-wave part of the spectrum of the material under study with a maximum of about 1.53 eV. This indicates the presence of biazafullerene (azafullerene dimers) in the composition of the obtained complex. Determining the belonging of the rest of this spectrum is the subject of our next work. Naturally, the structure of the energy spectrum of condensed matter, for obvious reasons, is richer than that of an individual molecule, and this is reflected in the corresponding emission spectra. But in addition to this, the more complex and broader spectrum in our case may also be due to the presence of other reaction products of $N_2$ and $C_{60}$ in the material under study. Along with the formation of azafullerene dimers, more complex structures of three or more molecules with a variety of spatial arrangements may be formed by the same principle of replacing carbon in the framework of the $C_{60}$ molecule with a nitrogen atom. Perhaps it is some of these formations that serve as quenching centers for the low-

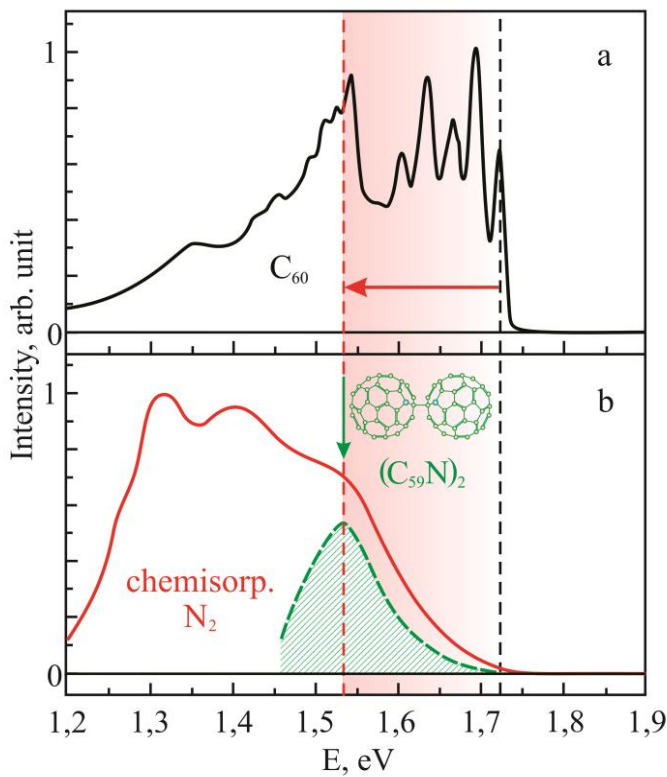

Fig. 2. The photoluminescence spectra of a polycrystalline $C_{60}$ samples at 20 K, excitation with $E_{exc}$ = 1.96 eV:

a - pure fullerite $C_{60}$ (solid black line);

b - $C_{60}$ saturated with molecular nitrogen at a temperature of 450 °C and a pressure of 30 atm - chemisorption (solid red line). The dashed green line (data from [12]) spectrum of a biazafullerene solution in toluene at normal conditions. The coincidence of the luminescence onsets with the maxima at 1.53 eV (810 nm) indicates the presence of azafullerene dimers in the studied samples. The spectra were normalized to their integral intensity and corrected for the spectral sensitivity of the setup.

temperature photoluminescence that we discovered in this substance earlier [6]. In addition, the question of the dependence of the diversity of the chemical composition of the complex under study on the sorption (reaction) temperature remains unclear. Apparently, even at such reaction temperatures (450 °C), the formation of cyanofullerenes with the general formula of their stable structures $C_{60}(CN)_{2n}$ (n=1, 2, 3, 4, …, 9) is quite likely, which are formed by a less energy-intensive mechanism of adding cyano groups. The degree of expression of acceptor properties in such compounds is directly dependent on the number of attached cyano groups. The presence of such substances in the studied samples, even in the form of impurities, can lead to both a noticeable complication of the crystal luminescence structure and to a non-uniform broadening of its bands. As in the case of hydrogen, we tried to record the primary chemical transformations in the immediate vicinity of the temperature of the onset of the chemical reaction – 420 °C. A further increase in the temperature of the reaction of $C_{60}$ with nitrogen leads to a very rapid increase in the lattice parameter of the matrix crystal [2], the nature of which may lie in an increase in the concentration of cyanofullerenes. As in the case of the formation of hydrofullerite $C_{60}H_X$, the addition of a cyano group outside the framework of the $C_{60}$ molecule to the released carbon bonds also increases the effective diameter of the modified molecule, but perhaps with an even more pronounced result. Perhaps this situation will become clearer after studying the features of the photoluminescence spectra of the reaction products at higher temperatures, which is planned in our future experiments on this topic.

**Hydrogen chemisorption.**

Fig. 3 shows the low-temperature (20 K) photoluminescence spectra of pure $C_{60}$ fullerite single crystals (black curve in Fig. 3a) and those saturated with hydrogen in the chemical sorption mode at 300 °C under 30 atm (blue curve in Fig. 3b), corrected for the spectral sensitivity of the setup and normalized to their integral intensity. Due to the normalization of curves a and b, Fig. 3 does not reflect the experimentally observed increase in the intensity of low-temperature photoluminescence of saturated fullerite relative to pure fullerite. The photoluminescence spectrum of $C_{60}$ after reaction with hydrogen is a sequence of inhomogeneously broadened bands, noticeably shifted toward higher energies. For the first maximum on the short-wave side of this spectrum and the luminescence edge of pure fullerite, the energy difference is 0.2 eV. This "blue" shift is shown in Fig. 3 by the arrow and indicates a widening of the HOMO - LUMO gap for chemically modified fullerite molecules in the crystal field of the lattice. Annealing of the samples in a dynamic vacuum (P = $10^{-3}$ mm Hg, T = 500 °C) did not restore the original form of the photoluminescence spectra of pure fullerite $C_{60}$, which means that we observe the luminescence of new substances, most likely fulleranes - products of hydrogenation of fullerene molecules. Since the saturation temperature (hydrogenation reaction) does not greatly exceed the adsorption crossover temperature of 250 ºC [1], the chemical interaction of $H_2$ and $C_{60}$ is ensured mainly by the presence of the catalytic effect of the fullerite itself on the dissociation of $H_2$ molecules and it proceeds quite moderately. In addition to the rate of interaction of the components, the reaction temperature also affects the degree of hydrogen saturation (hydrogenation index X) of the synthesized hydrofullerene molecules. Depending on the selected temperature of direct hydrogenation from the gas phase, the samples most often contain a mixture of several successive components of the fullerane series [11], with a predominance of one of the stable configurations: $C_{60}H_2$, $C_{60}H_8$, $C_{60}H_{12}$, $C_{60}H_{18}$, $C_{60}H_{24}$, $C_{60}H_{36}$, $C_{60}H_{44}$, $C_{60}H_{52}$, $C_{60}H_{60}$. Most likely, the diffuse broadening of the spectral bands observed in our experiments is explained by the presence in the reaction products of several fractions of molecules with different degrees of hydrogenation, as well as their isomers. Moreover, the noticeable "blue" shift of the onset of luminescence qualitatively corresponds to the previously established [10, 11] correlation between the degree of hydrogenation and the width of the forbidden band in hydrofullerenes. Theoretical calculations [11] for hydrofullerenes with the general formula $C_{60}H_{2n}$, where $2n = X$ is the hydrogenation index and n = 1, 2, 3, …, 30, also showed an increase in the HOMO – LUMO gaps with increasing X. The photoluminescence spectra of two stable forms of hydrofullerite $C_{60}H_{18}$ and $C_{60}H_{36}$ dissolved in toluene showed a high-energy shift of the corresponding emission onsets relative to those for a solution of pure fullerene $C_{60}$ in toluene, which is practically directly proportional to the index X. Thus, the "blue" shift of the spectra for a solution of fullerane $C_{60}H_{18}$ was about 0.6 eV, and for a solution of fullerane $C_{60}H_{36}$, it was, accordingly, twice as large, almost 1.2 eV. As stated above, in our experiment on crystalline samples a "blue" shift of the spectrum of only 0.2 eV was observed and, following the revealed proportionality, it should correspond to small hydrogenation indices X.

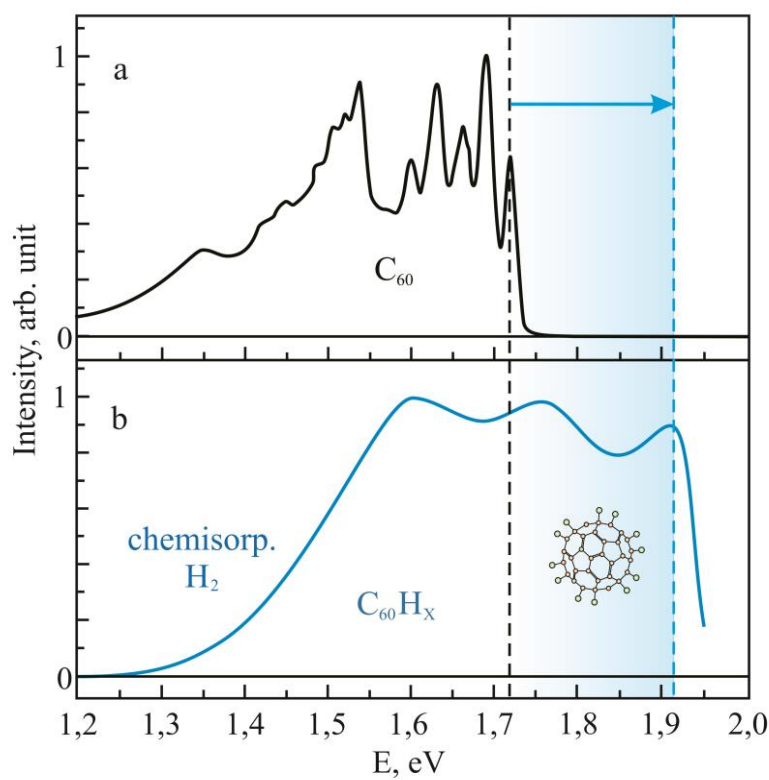

Fig. 3. The photoluminescence spectra of a polycrystalline $C_{60}$ samples at 20 K, excitation with a He–Ne laser ($E_{exc}$ = 1.96 eV):

a - pure fullerite $C_{60}$ (solid black line);

b - $C_{60}$ saturated with molecular hydrogen at 300 ºC and a pressure of 30 atm - chemisorption (solid blue line). The spectra were normalized to their integral intensity and corrected for the spectral sensitivity of the setup.

However, it is impossible to use the experimental and calculated data of the HOMO - LUMO gaps for solutions or isolated molecules in relation to the experimental data for crystals due to the vapor-crystal shift of the electron energy spectrum when the substance passes into the condensed state. For crystalline fulleranes, as for many other molecular solids, this shift occurs toward lower energies, therefore the calculated values of the band gap width for an isolated molecule and for a solution always significantly exceed both the experimental and calculated data for the solid state. In the work [10] the "blue" shift of the spectrum for crystalline $C_{60}H_{36}$ relative to pure $C_{60}$ was determined experimentally and is about 0.9 eV, which is less than 1.2 eV for solutions [11] and corresponds to the expected result. If we accept the practically direct proportionality of the spectrum shift and the width of the HOMO - LUMO gaps for solutions (in the region of small X) as valid for crystals, then the estimated value of the minimum hydrogenation index in our samples will be 8 or 10. However, it must be recognized that the strong inhomogeneous broadening of the spectral bands can be associated both with the presence of additional energy levels belonging to structural defects, and with the presence of other forms of hydrofullerites and their isomers, as already assumed above. Since the sensitivity of our recording equipment is reduced in the region of the onset of luminescence of the samples, we admit the presence of a larger "blue" shift than is actually observed. In addition, theoretical calculations [11] show close values of the HOMO – LUMO gaps in the range of hydrogenation indices X from 10 to 16 and therefore the real composition of the hydrogenation products in our case will presumably belong to the initial segment of the fullerane series with indices $8 \leq X \leq 14$.

These results show that the modification of $C_{60}$ by solid-phase hydrogenation is a promising method for changing the band gap of $C_{60}$ with acceptable control for obtaining functional luminescent materials with predetermined optical properties.

**Conclusions.**

The results of identification and analysis of photoluminescent emission spectra at low temperatures (20 K) of products of chemical sorption of nitrogen and hydrogen molecules from the gas phase by polycrystalline fullerene $C_{60}$ are presented for the first time. The formation of new chemical substances with their own radiative properties different from the luminescent properties of the initial material is recorded. In particular, opposite transformations of electron energy spectra and changes in quantum yield are established for cases of hydrogenation and nitridation of fullerite $C_{60}$. It is shown that in the first case, a mixture of hydrofullerites or fulleranes $C_{60}H_X$ with a close hydrogenation index is synthesized, with a spectrum shift to the high-energy region and a simultaneous increase in luminescence intensity. In the case of the reaction of $C_{60}$ with $N_2$, a multicomponent complex of nitrided fullerenes is formed, the overall spectrum of which is shifted to the long-wave region with pronounced low-temperature quenching of emission. Analysis of the "blue" shift of the hydrogenated fullerite spectrum showed that hydrogen chemisorption at temperatures close to the adsorption crossover temperature (250 °C) leads to the synthesis of weakly saturated fulleranes with a degree of hydrogenation X from 8 to 14. It was also found that the characteristic structureless luminescence of the azafullerene (biazafullerene) dimer $(C_{59}N)_2$ is present in the spectra of the reaction products of $C_{60}$ with $N_2$. Its maximum at 1.53 eV almost exactly coincides with the position of the first band in the spectrum of the synthesized compounds, and the "elongated" shape of its short-wave shoulder apparently makes the main contribution to the intensity of the beginning of the spectrum.

**Acknowledgments**

Authors thanks the National Academy of Sciences of Ukraine and the Ministry of Education and Science of Ukraine for their support.

**References**

[1] K.A. Yagotintsev, I.V. Legchenkova, Yu.E. Stetsenko, P.V. Zinoviev, V.N. Zoryansky, A.I. Prokhvatilov, and M.A. Strzhemechny, Saturation of fullerite $C_{60}$ with hydrogen: Adsorption crossover studies, Low Temp. Phys. 38 (2012) 952–956. DOI: 10.1063/1.4758781

[2] I.V. Legchenkova, K.A. Yagotintsev, N.N. Galtsov, V.V. Meleshko, Yu.E. Stetsenko, and A.I. Prokhvatilov, Effect of nitrogen sorption mechanisms on the properties of fullerite $C_{60}$ over a wide range of temperatures, Low Temp. Phys. 40 (2014) 685-689. DOI: 10.1063/1.4894316

[3] P.V. Zinoviev, V.N. Zoryansky, N.B. Silaeva, Yu.E. Stetsenko, M.A. Strzhemechny, and K.A. Yagotintsev, Orientational glassification in fullerite $C_{60}$ saturated with $H_2$: Photoluminescence studies, Low Temp. Phys. 38 (2012) 732-739. DOI: 10.1063/1.4746795

[4] Yu.E. Stetsenko, I.V. Legchenkova, K.A. Yagotintsev, A.I. Prokhvatilov, and M.A. Strzhemechny1, Intercalation of $C_{60}$ fullerite with helium and argon at normal temperature and pressure, Low Temp. Phys. 29 (2003)] 445-448. DOI: 10.1063/1.1542509

[5] G.E. Gadd, S. Moricca, S.J. Kennedy, M.M. Elcombe, P.J. Evans, M. Blackford., D. Cassidy, C.J. Howard, P. Prasad, J.V. Hanna, A. Burchwood, and D. Lavy, Novel rare gas interstitial fullerenes of $C_{60}$ with Ar, Kr and Xe, J. Phys. Chem. Solids 58 (1997) 1823-1832. DOI: 10.1016/S0022-3697(97)00096-6

[6] P.V. Zinoviev, V.N. Zoryansky, Photoluminescence of $C_{60}$ fullerite intercalated with nitrogen molecules wide range of temperatures, Low Temp. Phys. 48 (2022) 268-270. DOI: 10.1063/10.0009547

[7] I. Akimoto and K. Kan'no, Photoluminescence and Near-Edge Optical Absorption in the Low-Temperature Phase of Pristine $C_{60}$ Single Crystals, J. Phys. Soc. Jpn. 71 (2002) 630-643. DOI: 10.1143/JPSJ.71.630

[8] P.V. Zinoviev, V.N. Zoryansky, N.B. Silaeva, Photoluminescence of $C_{60}$ single crystals intercalated with molecular hydrogen, Low Temp. Phys. 34 (2008) 484-486. DOI: 10.1063/1.2920183

[9] P.V. Zinoviev, V.N. Zoryansky, V.V. Meleshko, Yu.E. Stetsenko, Effect of molecular nitrogen impurity on the photoluminescence of fullerite $C_{60}$, Low Temp. Phys. 41 (2015) 236-238. DOI: 10.1063/1.4915916

[10] K.P. Meletov, G.A. Kourouklis, High-pressure hydrogenated fullerenes: Optical spectra and stability of $C_{60}H_{36}$ at high pressure, J. Exp. Theor. Phys. 100 (2005) 760-774. DOI: 10.1134/1.1926437

[11] J.A. Teprovich, Jr., A.L. Washington, II, J. Dixon, P.A. Ward, J.H. Christian, B. Peters, J. Zhou, S. Giri, D.N. Sharp, J.A. Velten, R.N. Compton, P. Jenab, R. Zidan, Investigation of hydrogen induced fluorescence in $C_{60}$ and its potential use in luminescence down shifting applications, Nanoscale 44 (2016) 18760-18770. DOI: 10.1039/c6nr05998h

[12] B. Ma, A. Weitz, F. Wudl, The photophyiscal properties of azafullerenes $C_{59}HN$ and its dimer $(C_{59}N)_2$, Proc. Electrochem. Soc. 12 (1999) 359-367